\documentclass[journal,onecolumn,12pt]{IEEEtran}
\usepackage{epsf,epsfig,graphicx,pdfpages,amsmath,amssymb,hyperref}
\usepackage{subcaption} %\usepackage{subfigure}
\usepackage{wrapfig}
\usepackage{cite} %\usepackage{citesort}
\usepackage{tikz, pgfplots}
\usepackage[normalem]{ulem} % provides strikeout command \sout
\usetikzlibrary{arrows,positioning,snakes,shapes,decorations.markings} 
\usepackage{verbatim}
\graphicspath{{figs//}}

\usepackage{color}

\newcommand{\secref}[1]{Section~\ref{sec:#1}}
\newcommand{\Secref}[1]{Section~\ref{sec:#1}}

\newcommand{\Figref}[1]{Figure~\ref{fig:#1}}
\newcommand{\term}[1]{\textsl{#1}}
\newcommand{\Separate}{Separate-antenna}
\newcommand{\separate}{separate-antenna}
\newcommand{\Shared}{Shared-antenna}
\newcommand{\shared}{shared-antenna}
\newcommand{\Beff}{\text{\sf ENOB}}

\newcommand{\basestation}{base station}

\title{In-Band Full-Duplex Wireless: Challenges and Opportunities}
\author{Ashutosh Sabharwal, Philip Schniter, Dongning Guo, Daniel W.\ Bliss, Sampath Rangarajan, and Risto Wichman% 
\thanks{Ashutosh Sabharwal is at Rice University (partially supported by NSF CNS-1314822 and NSF CNS-1161596), 
Philip Schniter is at The Ohio State University (partially supported by the United States Air Force under Air Force contract FA8721-05-C-0002), 
Dongning Guo is at Northwestern University (partially supported by NSF ECCS-1231828), 
Daniel W.\ Bliss is at Arizona State University, 
Sampath Rangarajan is at NEC Laboratories America, Inc., and 
Risto Wichman is at Aalto University.} } 
\date{}
\begin{document}
\maketitle

% !TEX root = FD-tutorial.tex
\begin{abstract}
In-band full-duplex (IBFD) operation has emerged as an attractive solution for increasing the throughput of wireless communication systems and networks.  
With IBFD, a wireless terminal is allowed to transmit and receive simultaneously in the same frequency band.
This tutorial paper reviews the main concepts of IBFD wireless.
Because one the biggest practical impediments to IBFD operation is the presence of self-interference, i.e., the interference caused by an IBFD node's own transmissions to its desired receptions,
this tutorial surveys a wide range of IBFD self-interference mitigation techniques.
Also discussed are numerous other research challenges and opportunities in the design and analysis of IBFD wireless systems.
\end{abstract}

% !TEX root = FD-tutorial.tex
\section{Introduction}	\label{sec:intro}

The wireless revolution has resulted in ever-increasing demands on our limited wireless spectrum, driving the quest for systems with higher spectral efficiency.
Among the various ways to increase spectral efficiency, in-band full-duplex (IBFD) operation has recently gained attention. 
(See Section~\ref{sec:FD-papers} for a rapidly growing list of publications.)
The main idea behind in-band full-duplex is as follows.
Most contemporary communication systems contain terminals (e.g., \basestation{}s, relays, or mobiles) that function as both transmitters and receivers.
Conventionally, these terminals operate in half-duplex or out-of-band full-duplex, meaning that they transmit and receive either at different times, or over different frequency bands. 
Enabling wireless terminals to transmit and receive simultaneously over the same frequency band (i.e., IBFD operation) offers the potential to double their spectral efficiency, as measured by the number of information bits reliably communicated per second per Hz, and thus is of great interest for next-generation wireless networks.

Beyond spectral efficiency, full-duplex concepts can also be advantageously used beyond the physical layer, such as at the access layer. 
From the access-layer point of view, enabling \emph{frame
  level} in-band full-duplex, where a terminal is able to reliably receive an incoming frame while simultaneously transmitting an outgoing frame, could provide terminals with new capabilities. 
For example, terminals could detect collisions while transmitting in a contention-based network or receive instantaneous feedback from other terminals.

\begin{figure}
\centering
\scalebox{1.0}{\input{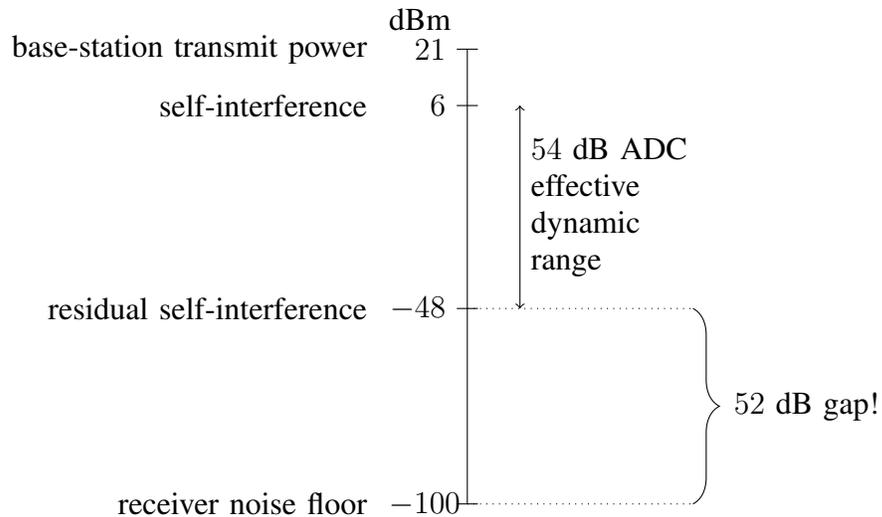}}
\caption{\small Illustrative example of residual self-interference motivated by contemporary femto-cell cellular systems \cite{SmallCell}. 
Even with perfect digital-domain self-interference cancellation, the effects of limited ADC dynamic range cause a residual self-interference floor that is $52$~dB above the desired receiver noise floor (i.e., the noise floor experienced by an otherwise equivalent half-duplex system).
}
\label{fig:CellularSNR}
\end{figure}

IBFD has, until now, not seen widespread use due to the potential debilitating effects of self-interference. 
Self-interference refers to the interference that a transmitting IBFD terminal causes to itself, which interferes with the desired signal being received by that terminal.
To appreciate the impact of self-interference, consider the following example in the context of contemporary femto-cell cellular systems \cite{SmallCell}. 
Based on data provided in~\cite[Table 10-2]{SmallCell}, femto \basestation{}s and mobile handsets transmit at $21$~dBm with a receiver noise floor of $-100$ dBm. 
If we assume $15$~dB isolation\footnote{Larger isolation is possible for different antenna architectures, e.g., see the experimental results reported in~\cite{Duarte10-FD-Feasibility,Choi10-Single-Channel-FD,Khojastepour11-FD-Cancellation,Everett11-Directional-Antenna-FD,Everett13-Passive}.} between the \basestation's transmit and receive signal paths, then the \basestation's self-interference will be $21-15 - (-100)=106$~dB above the noise floor. 
Thus, for a full-duplex base-station to achieve the link SNR equal to that of a half-duplex counterpart, it must suppress self-interference by more than $106$~dB---a daunting amount. (See \Figref{CellularSNR}.)
Moreover, the preceding analysis considers only femto-cell systems; systems with larger cells will require a higher transmit powers, and thus more self-interference suppression.

One might immediately wonder why an IBFD terminal cannot simply cancel self-interference by subtracting its transmitted signal from its received signal, given that the transmitted signal is known.
To start, if one attempts this cancellation in the digital domain, i.e., after the receiver's analog-to-digital converter (ADC), 
then ADC dynamic range is a major bottleneck for the following reasons.\footnote{
In practice, limited ADC dynamic-range is one of many factors that limit performance.  
Others include oscillator phase-noise, nonlinearities associated with amplifiers and mixers, IQ mismatch (if direct conversion is employed), and channel estimation error. 
}
Suppose that the IBFD terminal uses a $B$-bit ADC whose effective number of bits is $\Beff$.
Although we may be tempted to use the $6.02 \Beff$~dB rule-of-thumb to compute the resulting quantization error floor, this rule-of-thumb holds only in the absence of clipping, for which we need to budget an additional headroom of $\approx 6$~dB or $1$~bit (depending on the received peak-to-average-power-ratio).
Moreover, to prevent the system from being quantization limited, it is typical to place the quantization-error floor $\approx 6$~dB below the noise\footnote{Here, ``noise'' refers to all other additive impairments beyond self-interference and quantization error.} floor, which consumes $1$~additional bit.
Taking all of these factors into account, the \emph{effective} dynamic range of an ADC is closer to $6.02(\Beff-2)$~dB.
(See the Appendix for a more precise analysis.)
Thus, even if the self-interference is somehow \emph{perfectly} cancelled in the digital domain, residual errors due to quantization and noise will remain at about $6.02(\Beff-2)$~dB below the level of the self-interference at the ADC input.
So, if the IBFD terminal uses a $14$-bit ADC with $\Beff=11$~bits,\footnote{We note that the difference between the actual and effective number of bits is ADC dependent and generally grows with the operating bandwidth. (See for example~\cite{Walden99_ADC,Angeletti05_Evolution} for a more detailed discussion.)  Our example is based on high-end Analog Devices ADCs, and specifically the $14$-bit AD~9683~\cite{AD9683} designed for communications applications, which has an ENOB of $\approx11$ bits over its designed range of sampling bandwidths.} 
thus yielding an effective dynamic range of $6.02(11-2)\approx54$~dB, and if the self-interference power is $6$~dBm (as in the above femto-cell example), then a residual interference floor will remain at $6-54=-48$~dBm, which is $-48-(-100)=52$~dB above the desired femto-cell receiver-noise-floor of $-100$~dBm.
(See Fig.~\ref{fig:CellularSNR}). 

From the above example, it is evident that digital-domain cancellation can suppress self-interference only up to the \emph{effective} dynamic range of the ADC, i.e., 
$\approx 6.02(\Beff-2)$~dB.
This limitation is a serious one, since industry trends over the last few decades suggest that commercial ADCs have improved significantly in sampling frequency but only marginally in $\Beff$, improving by only about $1$~bit/decade~\cite{Angeletti05_Evolution} (see also~\cite{Walden99_ADC,Corcoran:2007aa}). 
We are thus strongly motivated to look for ways to reduce self-interference \emph{before} the ADC. 
Scanning the literature on self-interference cancellation, we observe that the proposed techniques use a combination of wireless-propagation-domain techniques \cite{Beasley90-FM-Radar,Cryan96-Active-Full-Duplex-Antenna,Chen98-Division-Free-Duplex,Anderson04AntennaIsoloation,Kim04-Cancellation-in-Radar,Kim06-Passive-Circulator-RFID,Kim2007RadarCanceller,BlissSSP07,Sangiamwong-FD-MUMIMO-Relay,Riihonen:2009aa,Larsson09-FD-MIMO-Relay,Hua:2010aa,Riihonen:2010aa,Choi10-Single-Channel-FD,Duarte10-FD-Feasibility,Duarte11-FD-Experimental-Characterization,Jain11-Real-Time-FD,Khandani12-FD-PPT,Aryafar12-MIDU,Duarte12_Wifi,Sahai11-Real-Time-FD,Khojastepour11-FD-Cancellation,Knox12-SingleAntenna,Bharadia13_fullduplex,Chun09-Self-Interference-Suppression-Relays,Lioliou10-FD-MIMO-Relay,Chun10Self-InterferenceNull,Senaratne:2011aa,Riihonen11FDMIMO,Snow11-TxRx-Beamforming-FD,Riihonen11-Optimal-Beamforming-FD-Relay,Everett11-Directional-Antenna-FD,Everett12-MastersThesis,Everett13-Passive,Day12FDMIMO,Day12FDRelay}, analog-circuit-domain techniques \cite{OHara63RadarNulling,Stove92-FMCW-Radar,Chen98-Division-Free-Duplex,suzuki99,Lin06RadarCanceller,Jun08LeakageCanceller,Lasser09-Broadband-Suppression-Leaking-Carriers,Pursula09RFID-Canceller,Duarte10-FD-Feasibility,Duarte11-FD-Experimental-Characterization,Sahai11-Real-Time-FD,Jain11-Real-Time-FD,Duarte12_Wifi,Bharadia13_fullduplex} and/or
digital-domain techniques \cite{suzuki99,Hamazumi00-RepeaterCanceller,Lin04FDRadar,BlissSSP07,Sangiamwong-FD-MUMIMO-Relay,Riihonen:2009aa,Ma09-FD-Repeater,Riihonen:2010aa,Duarte10-FD-Feasibility,Duarte11-FD-Experimental-Characterization,Sahai11-Real-Time-FD,Bharadia13_fullduplex,Choi10-Single-Channel-FD,Jain11-Real-Time-FD,Riihonen11FDMIMO,Senaratne:2011aa,RodriguezSPAWC13,RodriguezICASSP13,Day12FDMIMO,Day12FDRelay}.  
Moreover, each of these techniques may 
actively or passively suppress the self-interference contributed by device-extrinsic scattering effects.  (See the detailed discussion in \secref{techniques}.)

For example, in \term{wireless-propagation-domain} suppression, one aims to make the signal impinging upon the IBFD antennas free of self-interference.
Typically this is accomplished by some combination of antenna directionality, cross-polarization, and transmit beamforming.
In altering the transmit and/or receive antenna patterns, however, one must be careful to not also suppress the desired (outgoing and/or incoming) signal.
This latter concern motivates \term{analog-circuit-domain} cancellation methods, which tap a copy of the transmitted signal from an appropriate location in the transmitter and subtract it (after proper gain, phase, and delay adjustment) from each receive antenna feed, since doing so leaves the (transmit and receive) propagation patterns uncompromised.

Although judicious application of the aforementioned techniques can yield impressive self-interference suppression in an anechoic chamber, performance can degrade considerably due to environmental effects, such as nearby reflections, that are impossible to predict at the time of system design.
In essence, dealing with these device-extrinsic environmental effects requires learning and exploiting \term{channel-state information} (CSI), which is most easily handled in the \term{digital domain}.
Once it is available, CSI can be exploited not only for self-interference reduction, but also for \term{resource allocation}, e.g., the optimal allocation of limited signal power across space, time, and bandwidth \cite{BlissSSP07,Riihonen09-FD-Relay-Feasibility,Ju09FD_Resource,Ng10Hybrid_FD_HD,Riihonen11-Hybrid-FD-HD,Day12FDMIMO,Day12FDRelay}.

{\bf Why is IBFD important now?} 
As evidenced by the recent literature discussed in Section~\ref{sec:FD-papers}, there is significant interest in re-architecting terrestrial communications systems, such as WiFi and cellular systems, to leverage IBFD. 
A natural question is then, ``Why now?,'' considering that some of the IBFD self-interference mitigation techniques have been known for a while (see Section~\ref{sec:history}).
One driver is the widespread consensus that most ``traditional'' approaches to increasing spectral efficiency (e.g., advances in modulation, coding, MIMO) have by now been exhausted, leaving system designers willing to try more non-traditional approaches.
But perhaps a bigger driver is the architectural progression towards short-range systems, such as small-cell systems and WiFi,
where the cell-edge path loss is less than that in traditional cellular systems, making the self-interference reduction problem much more manageable. 
This shift towards smaller cells, coupled with the fact that ``consumer-faced'' data networks dominate in prevalence and market-size over radars and repeaters (which have employed IBFD for a long time, as discussed in \secref{history}), 
has sparked a renewed interest in IBFD.

Recently, several national, academic  and corporate research laboratories \cite{BlissSSP07,Duarte10-FD-Feasibility,Choi10-Single-Channel-FD,Khojastepour11-FD-Cancellation,Khandani10-FD-Patent} have taken the important first step of experimentally demonstrating the feasibility of IBFD in small-scale wireless communications environments (e.g., WiFi), thereby broadening the scope of IBFD beyond the traditional radar/repeater domain.
Successfully applying IBFD to large-scale commercial wireless networks will, however, require wireless researchers to tackle many more challenges, since IBFD ``removes'' one of the basic constraints in the design of wireless networks, i.e., the half-duplex  constraint. 
As a result, many aspects of network design and management will need to be reconsidered and redesigned.
Even after these design challenges have been overcome, it still may be that not every node in the network is suitable for IBFD operation.
For example, IBFD with small-form-factor devices still remains notoriously difficult and has, thus far, eluded successful experimental demonstration.
In the case that sufficient self-interference reduction is not possible, an alternative is ``virtual full-duplex" signaling~\cite{Guo:2010aa, Zhang:2013aa, Zhang:20XXaa}, which uses rapid on-off signaling to enable full-duplex at the \emph{frame} level while using half-duplex at the physical layer, thereby achieving higher network throughput (see \secref{net}).
In summary, preliminary results have shown that IBFD has strong potential to increase the spectral efficiency of future wireless networks, although additional design innovations are essential before IBFD is used in operational networks.

The remainder of this article is structured as follows.  
In \Secref{history}, we review the history of IBFD wireless, including its longstanding use in radar systems.
In \Secref{opportunities}, we describe several opportunities for IBFD in current and future wireless communications systems.
In \Secref{techniques}, we detail a number of interference suppression techniques that enable IBFD, 
and in \Secref{research}, we outline important ongoing research topics in IBFD.
Finally, in \Secref{conclusion}, we conclude.

% !TEX root = FD-tutorial.tex
\section{Literature Review}	\label{sec:history}

In-band full-duplex wireless has a long history, and in fact the concept has been in use since at least the 1940s. 
In this section, we first discuss the history of IBFD in radar systems, and then we discuss the history of IBFD in wireless communication systems.  
Due to lack of space, we focus on providing an exhaustive (to the best of our knowledge) list of citations as a resource for researchers, rather than a detailed discussion of every citation.

\subsection{Full-duplex Radars \label{sec:radar}}

\begin{figure}[htbp]
\centering
\begin{subfigure}[b]{0.35\textwidth}
        \centering
                \includegraphics[height=1.1in]{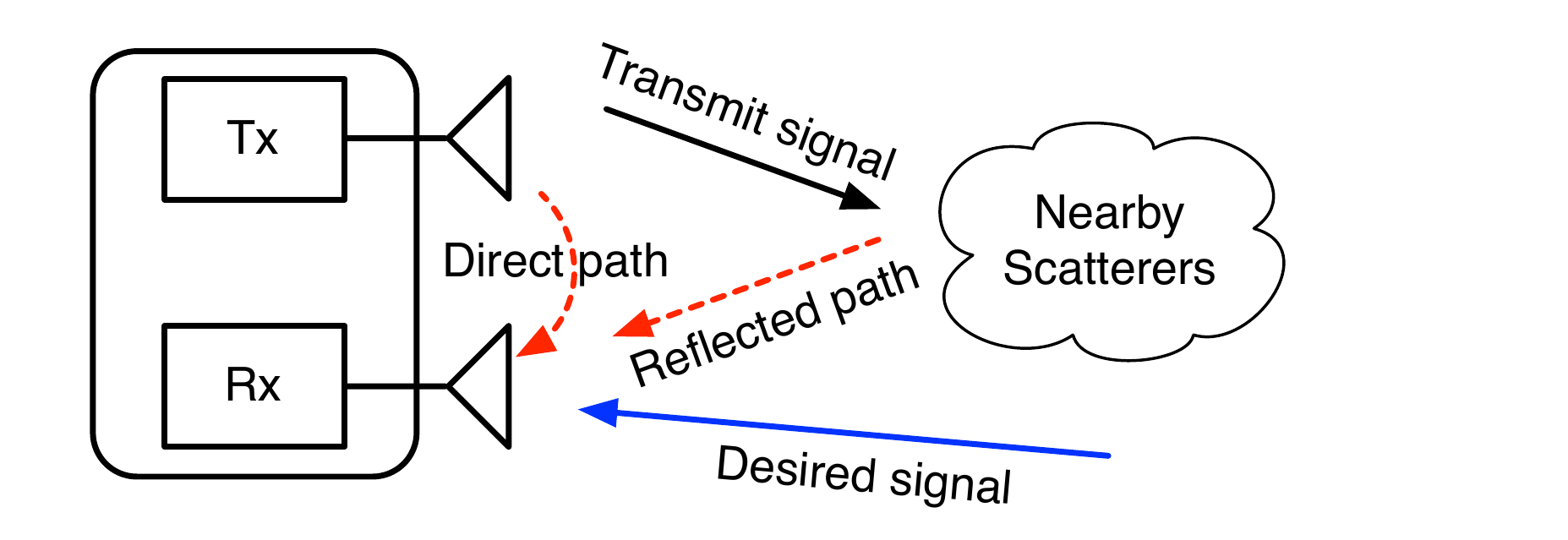}
        \caption{\Separate\ full-duplex \label{fig:separate-antenna}}
\end{subfigure}
\hspace{40pt}
\begin{subfigure}[b]{0.45\textwidth}
        \centering
       \includegraphics[height=1.1in]{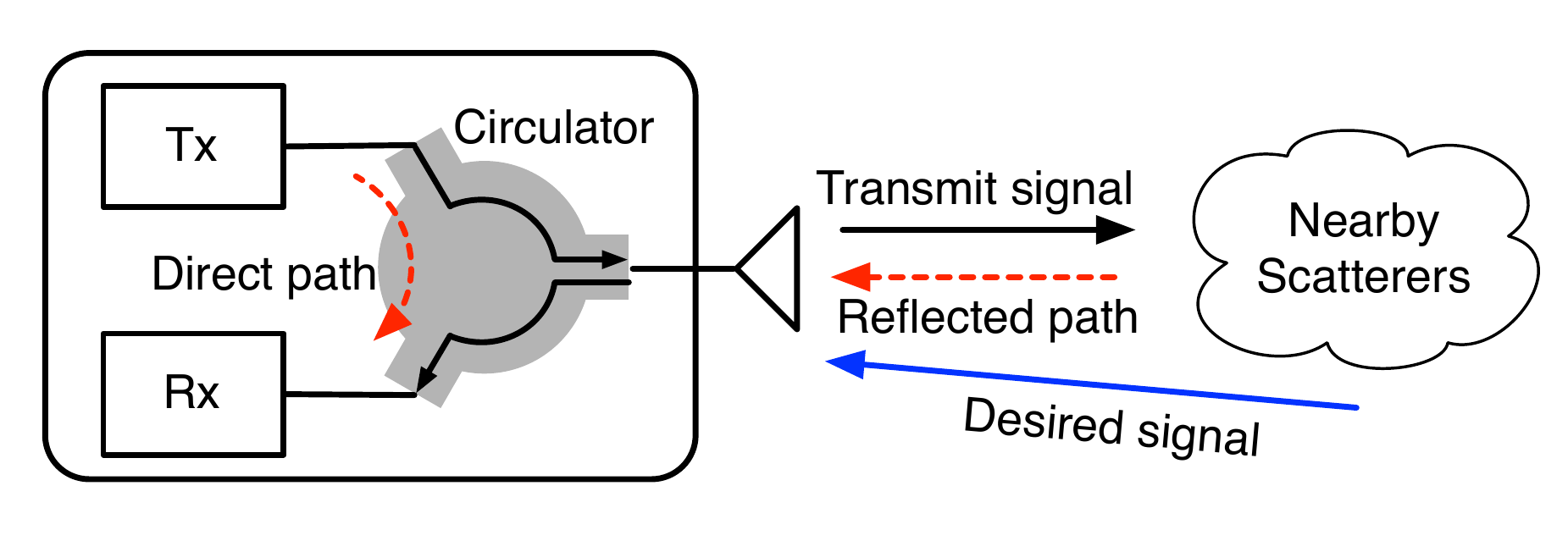}
        \caption{\Shared\ full-duplex \label{fig:shared-antenna}}
\end{subfigure}
\mbox{}\caption{\emph{Two methods of interfacing antennas to an IBFD wireless terminal.}
\label{fig:antenna-modes}}
\end{figure}

Continuous Wave (CW) radar systems use either two separate antennas (bistatic, as in \Figref{separate-antenna}), or one shared antenna (mono-static, as in \Figref{shared-antenna}), to transmit and receive simultaneously~\cite{Steer10-RF-Book}, in contrast to pulsed radar systems, which switch off the transmitter while radar returns are collected. 
Self-interference, often labeled as ``transmitter leakage'' in the radar literature~\cite{Stove92-FMCW-Radar,Lin04FDRadar}, is one of the key challenges faced in the design of all CW radars. 
The conventional CW radars of the 1940s and 1950s achieved isolation between the transmitter and receiver through antenna-separation-based path-loss in \separate\ systems, or through the use of circulators in \shared\ systems. 
(Circulators, illustrated in \Figref{shared-antenna} and discussed further in \secref{anatomy}, exploit nonlinear propagation in magnetic materials to isolate the incoming and outgoing signals \cite{WentworthEM}.)
Because only mild levels of isolation could be achieved using these techniques, keeping self-interference to a manageable level required strongly limiting the transmit power, which then strongly limited the detectable range  of targets. 
This restriction of CW radar to nearby targets (i.e., short ranges) turned out to be fortuitous, since detecting nearby targets with a pulsed radar system would require on/off switching times that are impractically small. Thus, the operational range of CW radars matched their need. 

In the 1960's, the ``feed-through nulling'' approach, an analog-circuit-based form of self-interference cancellation, was proposed to increase the dynamic range of CW radars~\cite{OHara63RadarNulling}. 
Although a total isolation of $60$~dB was reported, the canceler required a $60$+~kg block of precisely machined ferrite rotators and was very expensive. 
In 1990, an improved analog canceler was proposed in \cite{Beasley90-FM-Radar} that allowed adaptability to non-constant channel conditions, and its performance was demonstrated in \cite{Beasley91-Radar-Cancellation-Implementation}. 
Many refinements of the leakage canceler/isolator for monostatic CW radar have since been proposed~\cite{Lin04FDRadar, Lin06RadarCanceller,Kim04-Cancellation-in-Radar,Kim06-Passive-Circulator-RFID,Kim2007RadarCanceller,Pursula09RFID-Canceller,Jun08LeakageCanceller}. 
It is worth noting that CW radar technology forms the basis of the now-ubiquitous Gen-2 RFID protocol~\cite{RFID}; see also the articles~\cite{Pursula09RFID-Canceller,Lasser09-Broadband-Suppression-Leaking-Carriers} about ongoing research.

\subsection{Research Advances in In-band Full-duplex Wireless Communications\label{sec:FD-papers}}

While radar systems have extensively used IBFD, terrestrial wireless communication systems like cellular and WiFi have largely avoided IBFD, except in some special cases. 

The first known application of IBFD to wireless communication came in the 
context of \emph{relaying}, where in-band repeaters are used to increase coverage in a wireless communication system by receiving, amplifying, and re-transmitting the wireless signal in the same frequency band. 
(See \secref{opportunities} for more details on the relay topology.)
Relays are often used in tunnels or difficult terrains, where laying wireline backhaul to each terminal  is challenging and/or cost-prohibitive.  
For IBFD relays, the earliest known self-interference suppression techniques relied on increasing the physical separation between transmit and receive antennas (much like in bi-static CW radar)~\cite{Isberg89_Repeater_Performance,Slingsby92_CellEnhancer_Repeater,Slingsby95AntennaIsolation,Anderson04AntennaIsoloation}. 
Subsequently, a variety of circuit-domain self-interference cancellation strategies (including analog and digital techniques as well as adaptive and non-adaptive ones) were proposed, e.g.,
\cite{suzuki99,Hamazumi00-RepeaterCanceller,Kim03_Adaptive_FB_Cancellation_Repeater,Riihonen08Co-phasingFull-duplex,Riihonen08Co-phasingFilterFull-Duplex,Ma09-FD-Repeater,Riihonen11FDMIMO}. 
Further self-interference reductions were made possible through the use of 
antenna arrays, which allowed beamforming-based self-interference nulling~\cite{BlissSSP07,Larsson09-FD-MIMO-Relay,Sangiamwong-FD-MUMIMO-Relay,Lioliou10-FD-MIMO-Relay,Ju09-FD-Relays,Chun09-Self-Interference-Suppression-Relays,Riihonen11-Optimal-Beamforming-FD-Relay,Riihonen11FDMIMO,Chun10Self-InterferenceNull,Day12FDRelay}. 
Recently, a provision for IBFD relaying has been included in the 3GPP standard~\cite{3GPP-fullduplex}. 

The information-theoretic limits of IBFD relaying have also been studied.
Although early work~\cite{Shamai07-FD-AF-Relay,Cadambe08FullDuplex,Ju09FD_Resource,Simoens:2009aa,Sohaib:2009aa} focused on the ideal case of perfect self-interference cancellation, subsequent work~\cite{Riihonen09-FD-Relay-SINR,Yamamoto-FD-Relay,Song09-FD-Relay,Ju09-FD-Relays,Riihonen09-FD-Relay-Feasibility,Kan09FD-MIMO-thry,Baranwal12-Outage-Multihop-FD-Relays} included the effects of residual self-interference. 
More recently, information-theoretic analyses of MIMO IBFD relaying that include the effects of channel estimation error, finite ADC resolution, and amplifier nonlinearities on residual self-interference have been considered in~\cite{Day12FDRelay,BlissAWC2013}.

From the above body of work, it is clear that IBFD has a rich history in the context of wireless relays.
For other wireless system configurations, although a few early papers do exist (e.g., \cite{Chen98-Division-Free-Duplex}), the surge in interest has occurred only recently, after several academic/national/industrial research labs experimentally demonstrated the feasibility of \emph{bidirectional} IBFD~\cite{BlissSSP07,Radunovic09-FD-Indoor-Wireless,Duarte10-FD-Feasibility,Choi10-Single-Channel-FD,Khojastepour11-FD-Cancellation,Sahai11-Real-Time-FD,Jain11-Real-Time-FD,Khandani12-FD-PPT}, at least over short ranges.
In bidirectional IBFD, two modems simultaneously exchange messages over the same frequency band, as further described in \secref{opportunities}.
Much like CW radars, these implementations included both separate-antenna and shared-antenna architectures, as depicted in \Figref{antenna-modes}.
However, due to the somewhat limited levels of self-interference suppression demonstrated in these initial experiments, they did not convincingly establish the feasibility of IBFD for practical WiFi and cellular systems.
A second round of bidirectional IBFD experiments~\cite{Duarte11-FD-Experimental-Characterization,Duarte12_Wifi,Aryafar12-MIDU,Duarte12-Thesis}, however, demonstrated improved self-interference suppression, to levels sufficient for WiFi networks.
Further improvements were then experimentally demonstrated using advanced antenna designs in both separate-antenna~\cite{Everett11-Directional-Antenna-FD,Everett12-MastersThesis,Everett13-Passive} and shared-antenna~\cite{Knox12-SingleAntenna,Bharadia13_fullduplex} architectures. 
Moreover, several patents covering IBFD have been awarded \cite{Khandani10-FD-Patent, Fullerton97-FD-Ultrawide-Band-Communication,Stitzer82-Full-Duplex-Patent}.

While the aforementioned literature proposed and experimentally demonstrated successful bidirectional IBFD designs, a different body of recent work~\cite{Sahai12-PhaseNoise-Conference,Sahai12-PhaseNoiseJournal,BlissAsil12,Everett13-Passive,Ahmed:2013aa},\cite{Syrjala14_PhaseNoise-Analysis,Korpi14_ADC-Linearity-Challenges} has focused on characterizing the \emph{bottlenecks} of practical IBFD architectures through a combination of theory and experiments, with the goal of improving future system implementations.
Likewise, advanced self-interference cancellation solutions have been recently proposed~\cite{Day12FDMIMO,Ahmed:2013aa,Wichman13_Cancellation_Advanced,Bharadia13_fullduplex,Anttila13_Cancellation_PowerAmplifier,Choi13_Simultaneous-Transmission-Reception} that take into account the impairments introduced by the transmit radio chain. 
Moreover, information-theoretic analyses of MIMO bidirectional IBFD have been performed\cite{Day12FDMIMO} that include the effects of channel estimation error, finite ADC resolution, and amplifier nonlinearities on residual self-interference. 
Even in \emph{out-of-band} full-duplex transceivers, leakage due to transmit-chain impairments can cause self-interference at the receiver when the transmit and receive frequency bands are closely spaced. 
As a result, co-siting of TDD and FDD base stations operating on adjacent frequencies is not considered commercially viable~\cite{Holma:2004}.
The effects of such leakage, and associated cancellation algorithms, have been studied in~\cite{Kiayani14_Spurious-Emissions-in-FDD,Omer11_Nonlinear-Microwave-Components,Frotzscher12_Zeor-IF-Receiver,Kiayani13_TX-RX-Leakage}.

Beyond the relay and bidirectional topologies, progress has already been made towards understanding the capacity gains from IBFD in more general \emph{multi-user network topologies}, at least under perfect self-interference suppression~\cite{Cadambe08FullDuplex,Vaze12_DOF,Geng12_IAFD,Sahai13-FD-Uplink_Downlink}.
Similarly, distributed full-duplex~\cite{Bai12DecodeCancel,BaiTWC}, and the application of full-duplex to cognitive radios~\cite{Cheng11CongitiveFD} and multi-band systems~\cite{Schacherbauer01-Multiband-FrontEnd, Raghavan05-Interference-Canceller-Colocated-Radios}, have also been considered.

Furthermore, progress has also been made towards understanding the impacts of full-duplex beyond the physical layer, such as in full-duplex medium-access protocols~\cite{Singh11FDMAC,Sen11BackoffInFreq} and in cross-layer optimization for full-duplex network scheduling \cite{Weeraddana10-Self-Interference-Cancellation-Multihop,Ramirez13_Routing,Fang11-FD-MAC,Mesbah06-FD-MAC}.
Moreover, the advantages of full-duplex in neighbor discovery~\cite{Zhang:2013aa}, mutual broadcasting (e.g., for the exchange of network state information)~\cite{Zhang:20XXaa}, and ranging/localization~\cite{gan2013distributed} have also been considered. 

Based on the brief discussion above, it is clear that there exist many aspects of physical-layer IBFD and higher-layer full-duplex that require more thorough investigation.
We will describe a few of the important open challenges in Section~\ref{sec:research}.

% Local Variables:
% TeX-master: "FD-tutorial.tex"
% End:

% !TEX root = FD-tutorial.tex

\section{Opportunities to Leverage Full-duplex}	\label{sec:opportunities}

\begin{figure}[htbp]
\hfill
\begin{subfigure}[b]{1\textwidth}
	\centering
		\includegraphics[width=5in]{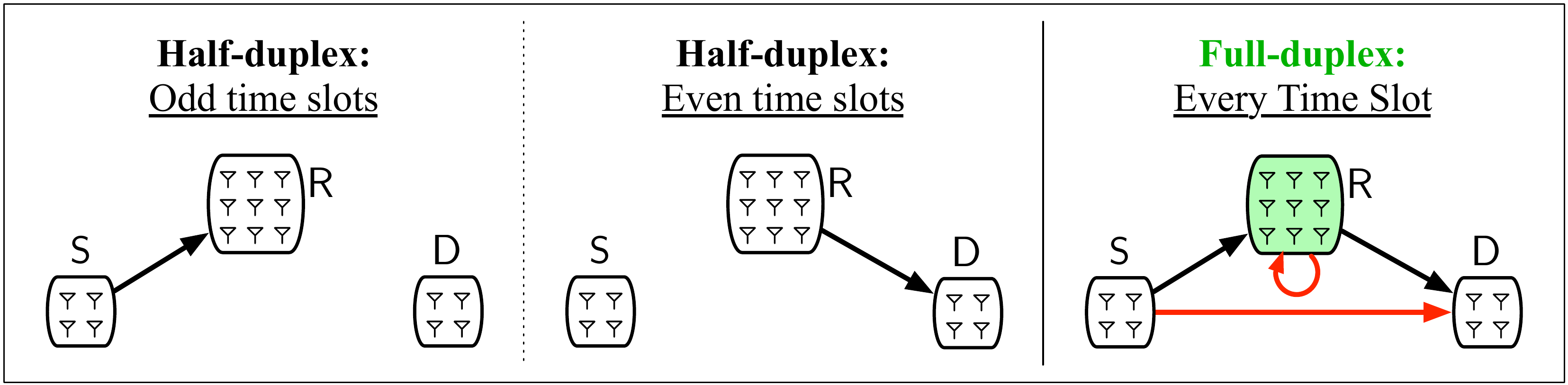}
 	\caption{Relay topology}
 	\label{fig:relay}
	\vspace{7pt}
\end{subfigure}
\begin{subfigure}[b]{1\textwidth}
	\centering
       \includegraphics[width=5in]{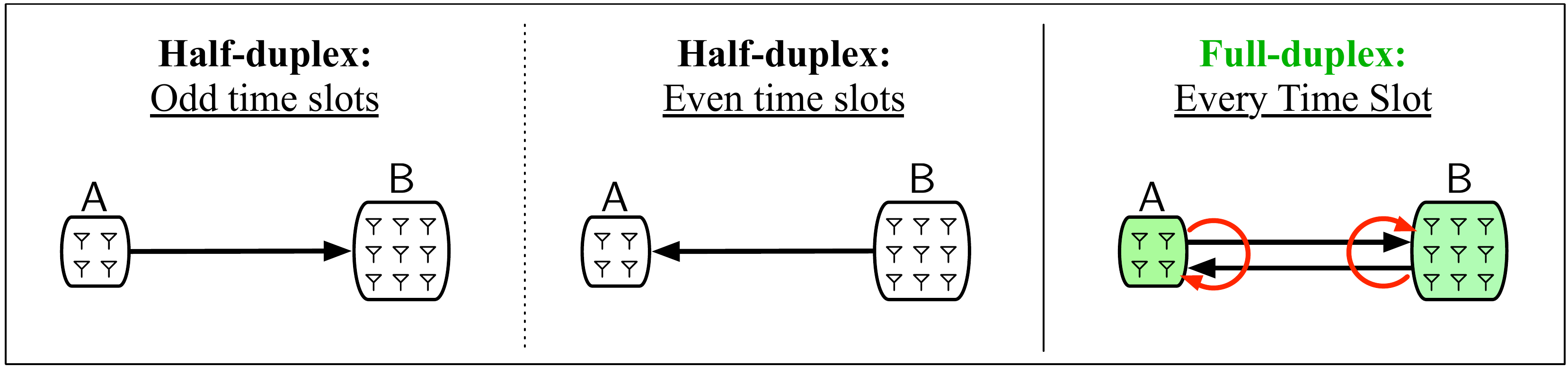}
 	\caption{Bidirectional topology}
 	\label{fig:bidirectional}
	\vspace{7pt}
\end{subfigure}
\begin{subfigure}[b]{1\textwidth}
	\centering
		\includegraphics[width=5in]{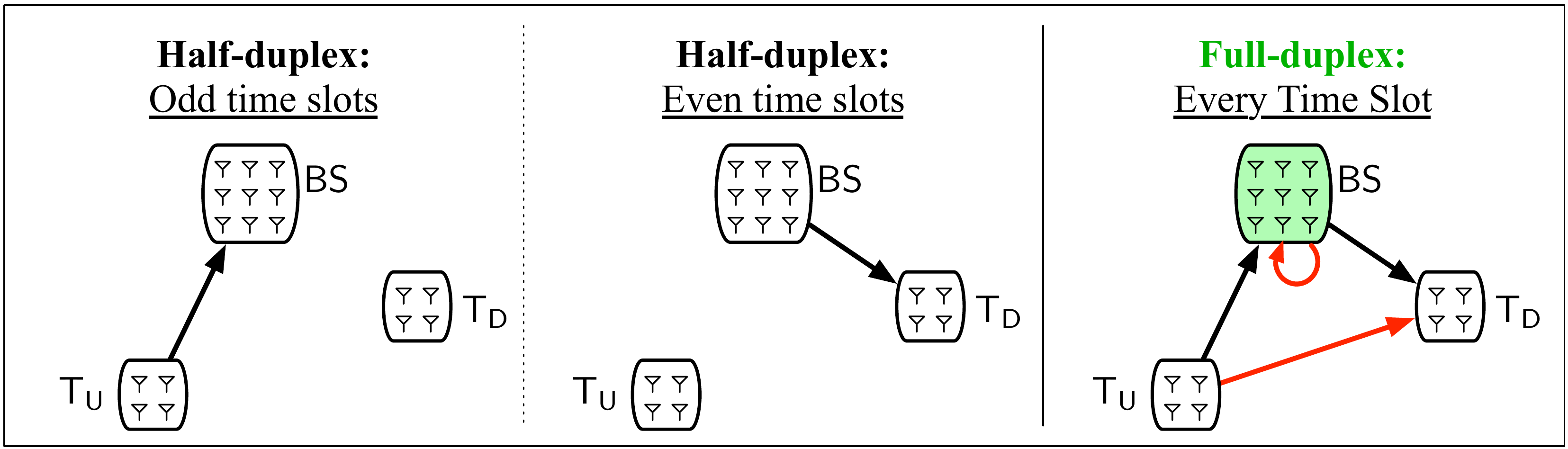}
 	\caption{\basestation{} topology}
 	\label{fig:three-node}
	\vspace{7pt}
\end{subfigure}
\hfill
\mbox{}
\caption{Three example topologies that illustrate the implications of IBFD terminals at the network level. IBFD-capable terminals are shaded green, and IBFD-induced interference is shaded red.
\label{fig:three-topologies}}
\end{figure}

To appreciate the gains from in-band full-duplex, it is instructive to look at three basic topologies in \Figref{three-topologies}: the (a)~relay topology, (b)~bidirectional  topology, and (c)~\basestation{} topology. 
In each case, the terminals can have multiple antennas and the number of antennas per terminal can differ across terminals. 
The three basic topologies are inspired by both ad~hoc and infrastructure-based (e.g., cellular and WiFi) networks. 
In an ad~hoc network with IBFD terminals, different traffic patterns can result in all three basic topologies manifesting in different parts of the network. 
Similarly, the bidirectional and \basestation{} topologies are possible in infrastructure-based networks with IBFD terminals.

First consider the \term{relay} topology shown in \Figref{relay}, where terminal~{\sf R} acts as a relay for the single flow of data being sent from source terminal~{\sf S} to destination terminal~{\sf D}.
If terminal~{\sf R} can only operate in half-duplex,\footnote{Here and in the other topologies, similar statements about spectral efficiency could be made if ``half-duplex'' was replaced by ``out-of-band full-duplex,'' and ``orthogonal time-slot'' was replaced by ``orthogonal frequency band.''} then it would need to alternate between receiving from terminal~{\sf S} and forwarding to terminal~{\sf D}, as shown in the left two panes of \Figref{relay}. 
However, if terminal~{\sf R} can operate in IBFD, then it could receive and forward simultaneously (over the same frequency band), as shown on the right pane of \Figref{relay}. 
Thus, by operating in IBFD mode, the relay network can potentially double\footnote{\label{foot:double}%
Whether IBFD truly doubles the spectral efficiency of half-duplex or not depends on the power-constraint assumed for the comparison. 
Under a \emph{peak} power constraint, the achievable spectral-efficiency of IBFD is exactly twice that of half-duplex (at all $\rm SNR$).
But under an \emph{average} power constraint, the IBFD system could use only half the power used by IBFD during its transmission slot, resulting in an achievable spectral-efficiency gain of strictly less than $2$ for any finite $\rm SNR$ \cite{Day12FDMIMO,Day12FDRelay}.
For example, if the SISO AWGN-channel capacity of the half-duplex system was $\log_2 (1+\mathrm{SNR})$ then, under equal average power and perfect self-interference suppression, the corresponding capacity for IBFD would be $2\log_2 (1+ {\rm SNR}/2)$, yielding a spectral-efficiency gain of $2\log_2 (1+\mathrm{SNR}/2)/\log_2 (1+\mathrm{SNR})$, which increases monotonically from $1\rightarrow 2$ as the $\rm SNR$ increases from $0\rightarrow \infty$.
This behavior becomes intuitive after recalling that communication systems transition from being power-limited to bandwidth-limited as $\rm SNR$ increases from $0\rightarrow \infty$ and that power-limited systems benefit little from an increase in bandwidth \cite{tv05}, which is effectively what IBFD provides. 
} the spectral efficiency (measured in bits/second/Hz) relative to half-duplex operation.
Note that the IBFD relay network requires only that the relay operate in full-duplex; neither the source nor the destination terminal needs to simultaneously transmit and receive.

Next consider the \term{bidirectional} topology shown in \Figref{bidirectional}, where there are two data flows: terminal~{\sf A} sends data to terminal~{\sf B}, and terminal~{\sf B} sends data to terminal~{\sf A}.
If either terminal~{\sf A} or terminal~{\sf B} only can operate in half-duplex, then communication from {\sf A}$\rightarrow${\sf B} cannot occur simultaneously with communication from {\sf B}$\rightarrow${\sf A}, and the two communications must be performed over orthogonal time slots, as shown in the left two panes of \Figref{bidirectional}.
However, if terminal~{\sf A} and terminal~{\sf B} both support IBFD operation, then communication from {\sf A}$\rightarrow${\sf B} can occur simultaneously with communication from {\sf B}$\rightarrow${\sf A}, as shown in the right pane of \Figref{bidirectional}, thereby potentially doubling\footnote{The comments on the doubling of spectral efficiency from half-duplex to full-duplex in footnote \ref{foot:double} hold for arbitrary topologies.} the spectral efficiency relative to half-duplex. 

Finally, consider the \term{\basestation{}} topology of \Figref{three-node}, where there also exist two data flows: terminal~{$\mathsf{T_U}$} sends data on the uplink to base station terminal {\sf BS}, and terminal~{\sf BS} sends independent data on the downlink to terminal~{$\mathsf{T_D}$}.
If terminal~{\sf BS} can only operate in half-duplex, then it would need to alternate between receiving from terminal~{$\mathsf{T_U}$} in one time slot and transmitting to terminal~{\sf $\mathsf{T_D}$} in an orthogonal time slot, as shown in \Figref{three-node}. 
However, if terminal~{\sf BS} could operate in IBFD, then it could support simultaneous in-band uplink and downlink communication, potentially doubling\footnote{The comments on the doubling of spectral efficiency from half-duplex to full-duplex in footnote \ref{foot:double} hold for arbitrary topologies.} the spectral efficiency.
Like in the relay topology, only terminal~{\sf BS} needs to support IBFD; terminals~{{$\mathsf{T_U}$} and {$\mathsf{T_D}$} need not simultaneously transmit and receive.

In \Figref{three-topologies}, the black arrows represent desired-signal propagation and the red arrows depict interference, the latter of which can come in two flavors. 
One is self-interference, which occurs when the signal transmitted by an IBFD terminal interferes with the reception of the desired incoming signal at the same terminal.
The second is inter-terminal interference, which occurs in IBFD networks between terminals that may themselves be non-IBFD. 
For example, in the relay topology of \Figref{relay}, terminal~{\sf S}'s transmission may cause interference to terminal~{\sf D}'s reception,
while in the \basestation{} topology of \Figref{three-node}, terminal~{$\mathsf{T_U}$}'s transmission may cause interference to terminal~{\sf $\mathsf{T_D}$}'s reception; such inter-terminal interference does not arise in the half-duplex versions of these networks.
Achieving the ideal doubling of spectral efficiency in an IBFD network requires managing both self- and inter-terminal interference. 
Since self-interference is the limiting factor of practical IBFD networks (e.g., it is much much stronger than inter-terminal interference), we will focus our attention on self-interference. 

Finally, we note that generalizations of the three example topologies presented in \Figref{three-topologies} can be easily imagined.
For example, the relay topology could support two simultaneous data-flows, one from {\sf S}$\rightarrow${\sf D} and another from {\sf D}$\rightarrow${\sf S}, if terminals~{\sf S}, {\sf R}, and {\sf D} all supported IBFD.
Similarly, the three-terminal \basestation{} topology could support four simultaneous flows: {{$\mathsf{T_U}$}}$\rightarrow${\sf BS}, {\sf BS}$\rightarrow${{$\mathsf{T_U}$}}, {{$\mathsf{T_D}$}}$\rightarrow${\sf BS}, and {\sf BS}$\rightarrow${{$\mathsf{T_D}$}, if terminals~{{$\mathsf{T_U}$}}, ${\sf BS}$, and {{$\mathsf{T_D}$}} all supported IBFD.
Or, the \basestation{} topology could be extended beyond three terminals to a network with a single IBFD \basestation{} and $N>2$ mobiles, supporting up to $2N$ simultaneous flows (if all terminals supported IBFD).
Finally, the bidirectional topology could be extended to a network with $N>2$ IBFD terminals that supports up to $N(N-1)$ simultaneous data flows.

From the higher-layer perspective, the existence of IBFD-capable terminals removes one of the most fundamental constraints assumed in conventional protocol design: the (frame level) half-duplex constraint, and therefore expands the design space.
This expanded design space offers new opportunities in protocol design that have the potential to increase network throughput.
For instance, many existing network protocols require knowledge of the state of 
communicating terminals, such as their power, modulation format, code rate, ACK, and queue length (to name a few). 
With IBFD capability, these terminals could potentially exchange state information even while in the presence of other co-located simultaneous transmissions. 
The ability to multiplex control information alongside data transmission offers the possibility to reduce protocol overhead\footnote{As a concrete example, access-layer protocol \emph{overhead} in IEEE~802.11 has increased with each generation of standards, and exceeds 50\% for peak PHY rates in IEEE~802.11n~\cite{Magistretti:2011aa}.} and significantly improve access-layer throughput.

% Local Variables:
% TeX-master: "FD-tutorial.tex"
% End:

% !TEX root = FD-tutorial.tex

\section{Techniques for Self-interference Reduction	\label{sec:techniques}}

We now discuss several approaches to self-interference reduction for IBFD terminals.
We find it convenient to partition them into three classes: propagation-domain, analog-circuit-domain, and digital-domain approaches.
Another important distinction is whether they 
actively or passively mitigate the self-interference caused by device-extrinsic scattering.

\subsection{Anatomy of an In-band Full-Duplex Terminal} \label{sec:anatomy}

To facilitate the discussion of self-interference suppression techniques, we first discuss the anatomy of a prototypical IBFD terminal and its immediate operating environment. 
\Figref{Anatomy} presents a schematic of a \separate\ IBFD terminal with multiple transmit antennas and multiple receive antennas.
A \shared\ IBFD terminal would look similar, except that each transmit-receive antenna pair in \Figref{Anatomy} would be replaced by a circulator attached to a common antenna, as in \Figref{shared-antenna}.
Differences between \separate\ and \shared\ systems will be further discussed in the sequel.

\begin{figure}[h]
\centering
\includegraphics[width=0.97\columnwidth]{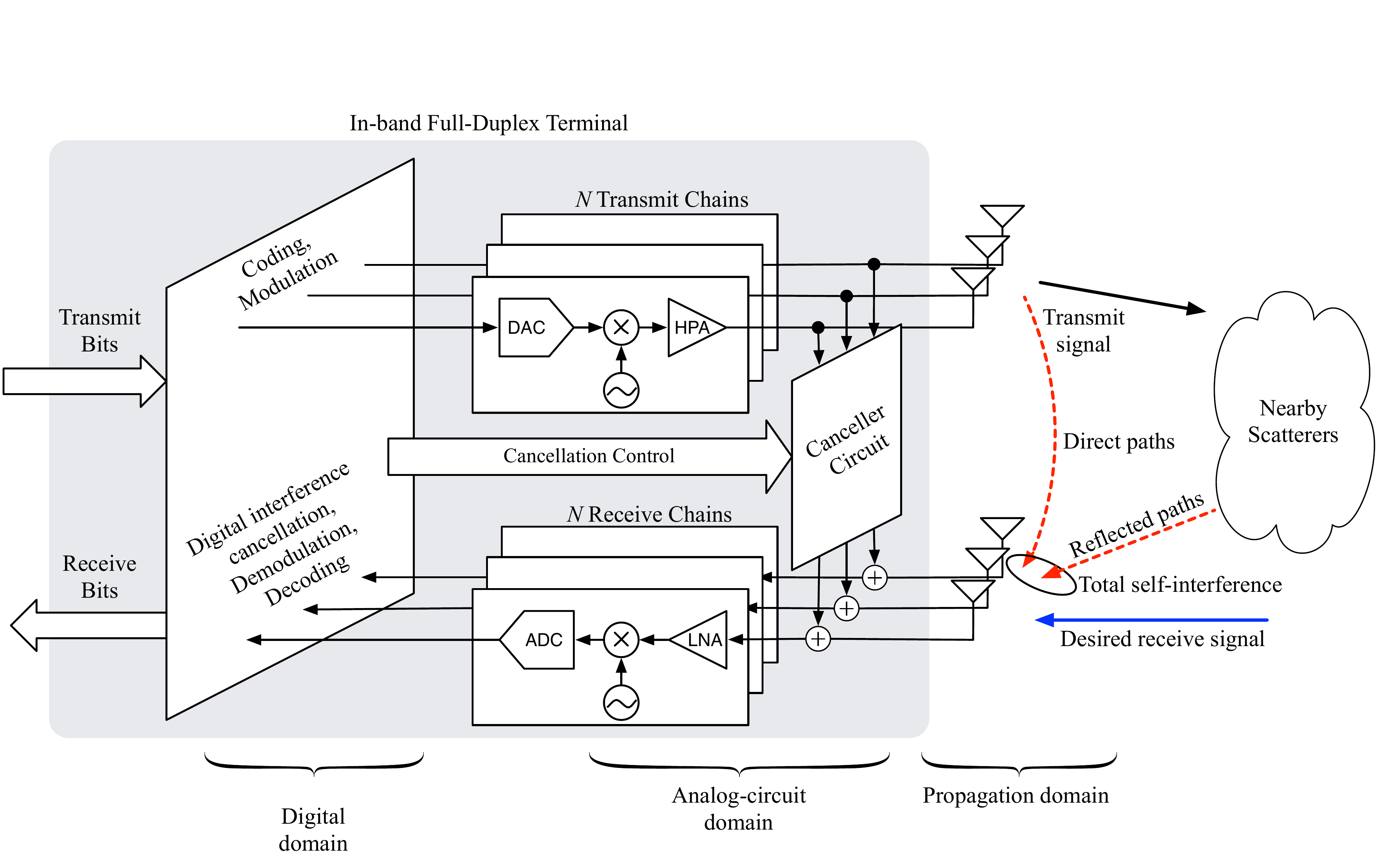}
\caption{Anatomy of a \separate\  in-band full-duplex terminal with multiple transmit antennas and multiple receive antennas.} 
\label{fig:Anatomy}
\end{figure}

As can be seen from \Figref{Anatomy}, the IBFD terminal accepts a transmit bitstream that is coded and modulated in the digital domain, producing a separate digital baseband signal for each transmit antenna.
Each of these digital signals is then converted to analog with a digital-to-analog converter (DAC), upconverted to a high carrier frequency, amplified using a high-power amplifier (HPA), and radiated using a transmit antenna.
In practice, this process will introduce several non-idealities into the transmit signal (e.g., DAC quantization noise, oscillator phase noise, and amplifier distortion), leading to small but important differences between the \emph{actual} and \emph{intended} transmit signals.

Simultaneously, and over the same frequency band, the IBFD terminal functions as a receiver. 
The signal picked up by each receive antenna is put through a separate hardware chain that includes a low-noise amplifier (LNA), downconverter, and analog-to-digital converter (ADC).
The resulting digital baseband signals are then jointly processed in the digital domain (involving demodulation, interference cancellation, and bit decoding) to produce the received bitstream.
Other aspects of the IBFD terminal, such as the canceler circuit visible in \Figref{Anatomy}, as well as other options not visible from the diagram, will be discussed in the sequel. 

There are two distinct ways in which the transmit and receive processing chains can be interfaced with antennas.
One is the \term{\separate}\ architecture, illustrated in both \Figref{separate-antenna} and \Figref{Anatomy}, and the other is the \term{\shared}\ architecture, illustrated in \Figref{shared-antenna}.
In the \separate\ architecture, each transmit chain uses a dedicated radiating antenna and each receive chain uses a dedicated sensing antenna \cite{Chen98-Division-Free-Duplex,Choi10-Single-Channel-FD,Duarte10-FD-Feasibility,Duarte11-FD-Experimental-Characterization,Jain11-Real-Time-FD,Aryafar12-MIDU,Duarte12_Wifi,Sahai11-Real-Time-FD,Khojastepour11-FD-Cancellation,Everett11-Directional-Antenna-FD,Everett12-MastersThesis,Everett13-Passive},
while in the \shared\ architecture, each transmit-chain/receive-chain pair shares a common antenna \cite{Beasley90-FM-Radar,Kim04-Cancellation-in-Radar,Kim06-Passive-Circulator-RFID,Knox12-SingleAntenna,Bharadia13_fullduplex}. 
\Shared\ operation requires a shared-antenna \term{duplexer} that routes the transmit signal from the transmitter to the antenna and routes the signal received on the antenna to the receiver,  all while isolating the receiver-chain from the transmit-chain. 
The most commonly used form of shared-antenna duplexer is the \term{circulator}, as illustrated in \Figref{shared-antenna}, which is a ferrite device that accomplishes the directional routing by exploiting non-linear propagation in magnetic materials \cite{WentworthEM}.  
Waveguide devices such as directional couplers~\cite{Kim06-Passive-Circulator-RFID} and coil-based devices such as hybrid transformers~\cite{Pursula08HybridTransformerRFID} can also be used to realize a shared-antenna duplexer. 
In the sequel, we use ``circulator'' when referring to any of these options.
An IBFD terminal with $N$ transmit chains and $N$ receive chains will require $2N$ antennas in a \separate\ implementation or $N$ antennas plus $N$ circulators in a \shared\ implementation. 

As important as what happens inside the IBFD terminal is what happens just outside the terminal.
We find it conceptually useful to decompose the received signal into three components: the \term{desired} received signal, the self-interference propagating \term{directly} from transmit chain to the receive chain, and the self-interference \term{reflecting} off device-extrinsic scatters, especially nearby ones. 
In \separate\ designs, the direct-path self-interference comprises the signal propagating directly from the IBFD terminal's transmit antennas to its own receive antennas, as shown in \Figref{separate-antenna},
whereas, in \shared\ designs, the direct-path self-interference is caused by circulator leakage (e.g., due to imperfect antenna matching), as shown in \Figref{shared-antenna}.
While direct-path self-interference can be accurately characterized offline (e.g., in an anechoic chamber) and thus addressed at system-design time, reflected-path self-interference cannot, since it depends on environmental effects that are changing and unpredictable.
Thus, practical IBFD terminals must use a combination of techniques to suppress both direct-path and reflected-path self-interference while maintaining strong (incoming and outgoing) desired-signal gain.

\subsection{Propagation-domain Self-interference Suppression} 
\label{sec:propagation}
Wireless-propagation-domain isolation techniques aim to electromagnetically isolate the transmit chain from the receive chain, i.e., to suppress the self-interference \emph{before} it manifests in the receive chain circuitry. 
The primary advantage to performing self-interference suppression in the propagation domain is that the downstream receiver hardware does not need to faithfully process signals with a huge dynamic range.
In \separate\ systems, propagation-domain isolation is accomplished using a combination of path loss \cite{Anderson04AntennaIsoloation,Duarte10-FD-Feasibility,Sahai11-Real-Time-FD,Khojastepour11-FD-Cancellation,Aryafar12-MIDU,Choi10-Single-Channel-FD}, cross-polarization \cite{Everett12-MastersThesis,Everett13-Passive,Khandani10-FD-Patent,Aryafar12-MIDU}, and antenna directionality \cite{Everett12-MastersThesis,Everett13-Passive}, while in \shared\ systems it is accomplished using a circulator.\footnote{In the case of the circulator, we consider the propagation domain to be the ferrite within the device.}

In \separate\ systems, the \term{path loss} between the IBFD terminal's transmit and receive antennas (or antenna arrays) can be increased by spacing them apart and/or by placing absorptive shielding between them, as quantified in \cite{Anderson04AntennaIsoloation,Duarte10-FD-Feasibility,Sahai11-Real-Time-FD,Everett12-MastersThesis,Everett13-Passive}.
Although path-loss-based techniques are attractive for reasons of simplicity, their effectiveness is greatly limited by the device form-factor: the smaller the device, the less room there is to implement such techniques.
\term{Cross-polarization} offers an additional mechanism to electromagnetically isolate the IBFD transmit and receive antennas.
For example, one may build an IBFD terminal that transmits only horizontally polarized signals and receives only vertically polarized signals with the goal of avoiding interference between them \cite{Khandani12-FD-PPT,Aryafar12-MIDU,Everett13-Passive}.
Similarly, with \term{directional} transmit and/or receive antennas (i.e., antennas with non-uniform radiation/sensing patterns), one may align their null directions in an attempt to achieve the same goal \cite{Everett11-Directional-Antenna-FD,Everett13-Passive}. 
In fact, one can build a highly directional transmit antenna from two omni-directional antennas using careful \term{antenna placement} \cite{Khandani10-FD-Patent,Choi10-Single-Channel-FD,Khojastepour11-FD-Cancellation,Aryafar12-MIDU}: by placing a single receive antenna at precisely a location where the carrier waveforms are exactly $180$~degrees out of phase, any narrowband signal modulated on those carriers will be near-perfectly canceled.

Although the preceding path-loss, cross-polarization, and antenna-directionality exploiting techniques are, in practice, somewhat hindered by issues like placement sensitivity and device integration, they can still be quite effective at suppressing \emph{direct-path} self-interference.
For example, using commercially available hardware, self-interference suppression levels of $74$~dB have been attained in anechoic-chamber settings \cite{Everett13-Passive}. 
Such anechoic figures are more representative of outdoor deployments than indoor ones, since the former are generally less reflective. 
In fact, when combined with analog cancellation (see \secref{analog}) and digital cancellation (see \secref{digital}), the design from~\cite{Everett13-Passive} allowed near-perfect outdoor IBFD over ranges up to $150$ meters.
The Achilles heel of these techniques, however, is their sensitivity to \emph{reflected-path} self-interference, whose channel characteristics are unknown at the time the system is designed.
For example, the same design that delivered $74$~dB suppression in the anechoic chamber delivered only $46$~dB suppression in highly reflective indoor office environments \cite{Everett13-Passive}. 
Note that essentially the same problems occur in \shared\ systems: 
a high-precision circulator (i.e., one with very low transmitter leakage) may suppress direct-path self-interference very well, but it has no way of discriminating between reflected-path self-interference and the desired receive-signal.

Handling device-extrinsic effects like nearby multipath scattering requires a \term{channel aware} method that can respond to these effects.
\term{Transmit beamforming} is an example of a channel-aware propagation-domain self-interference suppression strategy, for which the IBFD terminal's multi-antenna transmit array is electronically steered (through the adjustment of per-antenna complex weights) in an attempt to zero the radiation pattern at each of the IBFD receive antennas.
\cite{BlissSSP07,Chun09-Self-Interference-Suppression-Relays,Lioliou10-FD-MIMO-Relay,Chun10Self-InterferenceNull,Riihonen11FDMIMO,Snow11-TxRx-Beamforming-FD,Riihonen11-Optimal-Beamforming-FD-Relay}. 
Of course, doing this successfully in the presence of reflected self-interference requires accurate knowledge of the (direct and reflected) self-interference channel gains and delays, which may be learned either explicitly via channel-estimation or implicitly via weight adaptation.
Moreover, each null placement consumes one degree-of-freedom in the weight-vector design (beyond the constraint imposed by the transmit power constraint), and so effective transmit beamforming requires more IBFD transmit antennas than receive antennas. 
Finally, it may be worth noting that, while receive beamforming can also be performed, we would not classify it as propagation-domain interference suppression because the nulling would occur at the antenna-combining point, which resides in the circuit-domain.

Whether channel-aware or channel-unaware, the aforementioned propagation-domain self-interference suppression techniques have a potentially serious weakness: approaches that adjust the IBFD transmit and/or receive patterns to suppress self-interference might accidentally suppress the (incoming or outgoing) \emph{desired} signals as well.
This concern motivates the methods discussed in the sequel.

\subsection{Analog-circuit-domain Self-interference Cancellation}	\label{sec:analog}

Analog-circuit-domain cancellation techniques aim to suppress self-interference in the analog receive-chain circuitry, before the ADC. This suppression may occur either before or after the downconverter and the LNA.
\Figref{Anatomy} shows one such configuration, where the transmit signal is tapped at the transmit antenna feed, electronically processed in the analog-circuit domain, and subtracted from the receive-antenna feed in order to cancel self-interference.
But other options exist, such as tapping the transmit signal in the digital domain, applying the necessary gain/phase/delay adjustments digitally (where it is much easier to do so), and then converting\footnote{Converting to analog requires an additional DAC and upconverter (per transmit chain), but it does not require an additional high-power amplifier, which is one of the most power-hungry and expensive components in the transmit chain. We note that, while digital generation of the canceling signal can be implementationally more convenient, it does face imperfections due to DAC and up-converter phase noise, and thus this (as well as any) architectural choice needs to be carefully evaluated.} it to the analog-circuit domain for use in self-interference cancellation \cite{Duarte10-FD-Feasibility,Duarte11-FD-Experimental-Characterization,Duarte12_Wifi}, as depicted by the ``cancellation control'' path in \Figref{Anatomy}.
That said, tapping the outgoing signal as close as possible to the transmit antenna has advantages, since doing so better captures the presence of transmitter non-idealities like oscillator phase-noise and HPA distortion.
Likewise, placing the cancellation point as close as possible to the receive antenna has the advantage of freeing more downstream hardware from the need to faithfully process signals with high dynamic range.

Like their propagation-domain counterparts, analog-circuit-domain cancellation techniques can either be channel-aware or channel-unaware.
Channel-unaware techniques aim to cancel only the direct-path interference, whereas channel-aware techniques attempt to cancel both the direct- and reflected-path interference. 
The ``passive suppression'' techniques~\cite{Duarte10-FD-Feasibility,Everett11-Directional-Antenna-FD,Duarte11-FD-Experimental-Characterization,Duarte12_Wifi,Everett13-Passive} would be examples of channel-unaware analog-circuit-domain schemes, whereas the ``active cancellation'' techniques~\cite{Duarte10-FD-Feasibility,Choi10-Single-Channel-FD,Khojastepour11-FD-Cancellation,Jain11-Real-Time-FD,Bharadia13_fullduplex}  would be examples of channel-aware analog-circuit-domain schemes.

For sufficiently narrowband signals, the (direct and/or reflected) self-interference channel can be well modeled as a complex gain and delay between every transmit and receive antenna pair.
In this case, a \separate\ IBFD terminal with single transmit and receive antennas would need to adjust only a single scalar complex cancellation gain and a single delay.
A channel-unaware approach would do this adjustment once, when the system is designed/calibrated, while a channel-aware design would continually adjust this gain and delay to compensate for changes in the reflection channel. 
Many implementations of the canceler circuit (as in \Figref{Anatomy}) have been proposed; see e.g.,~\cite{Choi10-Single-Channel-FD,Duarte10-FD-Feasibility,Aryafar12-MIDU,Bharadia13_fullduplex} for implementations that leverage a different mix of analog and digital processing components. 
The extension to multiple IBFD transmit and/or receive antennas will be straightforward, and will require tracking a separate gain and delay for every transmit-receive antenna \emph{pair}.

For wideband signals, the direct-path self-interference can be mitigated using the same analog-circuit-domain techniques described above if the antenna gain and phase responses are engineered to be frequency-flat. 
The reflected-path self-interference channel, however, will in general be frequency-selective as a result of multipath propagation. 
This frequency selectivity makes analog-circuit-domain cancellation much more challenging, especially when the transmit signal is tapped and processed in the analog domain (as in \Figref{Anatomy}), since then it would require adapting an \emph{analog} filter for each transmit-receive antenna pair.
Tapping the transmit signal in the digital domain, however, facilitates the use of \emph{digital} adaptive filtering for reflected-path self-interference cancellation, which is typically much easier to implement, as proposed in~\cite{Duarte12_Wifi} for wideband OFDM signals. 

The preceding discussion demonstrates that there are many options and tradeoffs involved in the design of analog-circuit-domain interference cancellation.
In summary, approaches that tap and cancel the interfering transmit signal very close to the antennas have the advantage of circumventing analog-domain non-idealities like HPA distortion and phase-noise, but the disadvantage of requiring analog-domain signal processing, which becomes especially difficult in the case of wideband reflected-path self-interference.
Meanwhile, approaches that tap and process the interfering transmit signal in the digital domain, for subsequent use in analog-domain cancellation, have the advantage of facilitating sophisticated adaptive DSP techniques to reflected-path self-interference, but the disadvantage that their cancellation precision is limited by downstream analog-circuit non-idealities.

\subsection{Digital-domain Self-interference Cancellation}	\label{sec:digital}
Digital-domain cancellation techniques aim to cancel self-interference \emph{after} the ADC by applying sophisticated DSP techniques to the received signal.
The advantage of working in the digital domain is that sophisticated processing becomes relatively easy. 
For example, while receive beamforming (where the per-antenna received signals are weighted by separate adaptive complex-valued gains before being summed together) could in principle be implemented in the analog-circuit domain, it is far more common in practice to implement it digitally \cite{BlissSSP07,Sangiamwong-FD-MUMIMO-Relay,Lioliou10-FD-MIMO-Relay,Riihonen11-Optimal-Beamforming-FD-Relay}, for reasons of circuit complexity and power consumption.
As described in \Secref{intro}, however, the disadvantage to digital-domain cancellation is that the ADC's dynamic-range limits the amount of self-interference reduction that is possible.
Thus, for digital-domain methods to be successful, a sufficient amount of the self-interference suppression must first occur \emph{before} the ADC, using the propagation-domain and/or analog-circuit-domain methods described above.
In this sense, we can think of digital-domain cancellation as the \emph{last} line of defense against self-interference, where the goal is to cancel the self-interference left over from the propagation-domain and analog-circuit-domain approaches.

The first step to designing a digital-domain cancellation algorithm is to build a (baseband-equivalent) discrete-time system model that captures everything between the IBFD terminal's DAC and ADC, including the propagation-domain and analog-circuit-domain interference suppression techniques described earlier.
An accurate model would need to capture not only the IBFD terminal itself, but also the inter-antenna propagation channels, and possibly even the other terminals in the IBFD network (which themselves may or may not support IBFD operation; recall \Secref{opportunities}).
\Figref{Digital} shows examples of such models for the relay and bidirectional topologies, taken from \cite{Day12FDRelay,Day12FDMIMO}. 

In \Figref{Digital}, $\sqrt{\rho}\boldsymbol{H}$ and $\sqrt{\eta}\boldsymbol{H}$ are used to model signal and interference propagation, respectively, including the effects of propagation-domain and analog-circuit-domain suppression.
In the frequency-selective (i.e., wideband signal) case with multiple transmit and receive antennas, these quantities would be matrices of $z$-domain polynomials \cite{RodriguezSPAWC13,RodriguezICASSP13}, simplifying to matrices of complex numbers in the non-selective (i.e., narrowband signal) case, or scalar-valued entities in the case of single transmit and receive antennas. 
\Figref{Digital} also models the important difference between the intended and actual transmit signals, denoted by $\boldsymbol{x}$ and $\boldsymbol{s}$, respectively, through the use of a ``transmitter distortion'' signal $\boldsymbol{c}$ that may depend (nonlinearly) on $\boldsymbol{x}$. 
Recall that the intended and actual transmitted signals may differ as a result of DAC quantization noise~\cite{Riihonen:2012asi2}, upconverter phase-noise, and HPA distortion.
Similarly, \Figref{Digital} models the important difference between the received wireless signal and the corresponding ADC-output, denoted by $\boldsymbol{u}$ and $\boldsymbol{y}$, respectively, through the use of a ``receiver distortion'' signal $\boldsymbol{e}$ that may depend (nonlinearly) on $\boldsymbol{u}$.
Note that these transmitter and receiver distortions are independent of the additive channel noise $\boldsymbol{n}$.

\begin{figure}[h]
  \centering
  \begin{subfigure}[b]{0.50\textwidth}
    \centering
    \scalebox{0.80}{\input{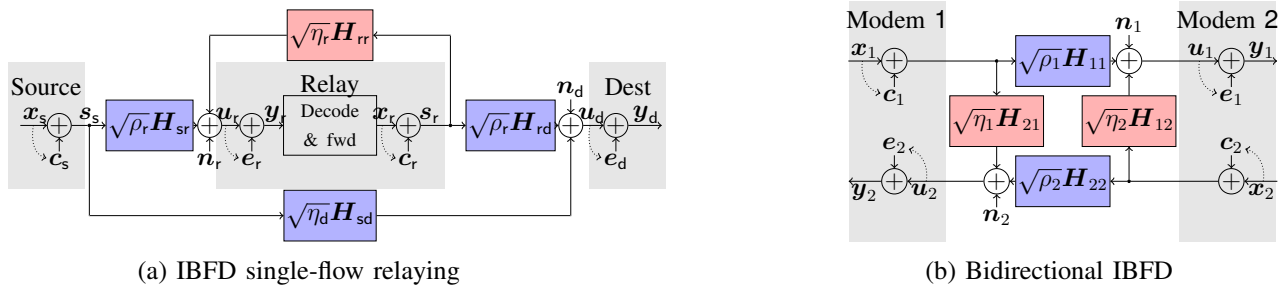}}
    \caption{IBFD single-flow relaying} 
    \label{fig:RelayDigital}
  \end{subfigure}
  \hfill
  \begin{subfigure}[b]{0.40\textwidth}
    \centering
	\scalebox{0.80}{\tikzstyle{block} = [draw,fill=red!30, minimum size=2em, inner sep=1pt]
\tikzstyle{sum} = [draw,shape=circle, minimum size=1em]
\tikzstyle{bigBlock} = [draw,fill=blue!50,minimum size=4em]
\tikzstyle{split} = [draw,shape=circle, fill=black,  inner sep=0pt, minimum size=.1em]
\tikz[decoration={
  markings,
  mark=at position 1.0 with {\arrow{>}}}
]{
\path[fill=black!10] (-.25,-1) rectangle ++(+1.60,4) ++ (-.8,-.3) node {Modem {\sf 1}};
\path[fill=black!10] (6.84,-1) rectangle ++(-1.60,4) ++ (+.8,-.3) node {Modem {\sf 2}};
%\path[fill=blue!10] (6,-1) rectangle (1.35,3) ++ (-.8,-.3) node {Modem {\sf A}};
%\path[fill=blue!10] (-.25,-.5) rectangle (1.3,2.5) ;
%
\path (0,2) node[anchor = south,inner sep=1pt] (x1) {$\boldsymbol{x}_1$}
++(.5,0) node[sum] (sum1) {} ++(0,0) node  {$+$}
++(1.7,0) node[split] (split1) {}
++(1.1,0) node[block, fill=blue!30] (H11) {$\sqrt{\rho_1}\boldsymbol{H}_{11}$}
++(1.1,0) node[sum] (sum2) {} ++(0,0) node  {$+$}
++(1.7,0) node[sum] (sum3) {} ++(0,0) node  {$+$}
+(-.5,0) node[anchor = south,inner sep=1pt] (u1) {$\boldsymbol{u}_1$}
++(.5,0) node[anchor = south,inner sep=1pt] (y1) {$\boldsymbol{y}_1$};
\path (sum1) +(0,-.6) node[inner sep=1pt] (c1) {$\boldsymbol{c}_1$};
\path (sum2) +(0,+.6) node[inner sep=1pt] (n1) {$\boldsymbol{n}_1$};
\path (sum3) +(0,-.6) node[inner sep=1pt] (e1) {$\boldsymbol{e}_1$};
\path (split1) ++(0,-1) node[block] (H21) {$\sqrt{\eta_1}\boldsymbol{H}_{21}$};
\path (sum2) ++(0,-1) node[block] (H12) {$\sqrt{\eta_2}\boldsymbol{H}_{12}$};
%\node (u1) [left of  = sum3] {$\boldsymbol{e}_1$};
%
%
%
\draw[->] (x1.south west) to (sum1);
\draw[->] (c1) to (sum1);
\draw[->,densely dotted, out = -100, in = 165] (x1) to (c1);
\draw (sum1) to (split1);
\draw[->] (split1) to (H11);
\draw[->] (H11) to (sum2);
\draw[->] (sum2) to (sum3);
\draw[->] (sum3) to (y1.south east);
\draw[->] (n1) to (sum2);
\draw[->] (e1) to (sum3);
\draw[->,densely dotted, out = -100, in = 165] (u1) to (e1);
\draw[->] (split1) to (H21);
\draw[->] (H12) to (sum2);
\path (0,0) node[anchor = north,inner sep=1pt] (y2) {$\boldsymbol{y}_2$}
++(.5,0) node[sum] (sum6) {} ++(0,0) node  {$+$}
+(.5,0) node[anchor = north,inner sep=1pt] (u2) {$\boldsymbol{u}_2$}
++(1.7,0) node[sum] (sum5) {} ++(0,0) node  {$+$}
++(1.1,0) node[block, fill=blue!30] (H22) {$\sqrt{\rho_2}\boldsymbol{H}_{22}$}
++(1.1,0) node[split] (split2) {} ++(0,0) node  {}
++(1.7,0) node[sum] (sum4) {} ++(0,0) node  {$+$}
++(.5,0) node[anchor = north,inner sep=1pt] (x2) {$\boldsymbol{x}_2$};
\path (sum4) +(0,+.6) node[inner sep=1pt] (c2) {$\boldsymbol{c}_2$};
\path (sum5) +(0,-.6) node[inner sep=1pt] (n2) {$\boldsymbol{n}_2$};
\path (sum6) +(0,+.6) node[inner sep=1pt] (e2) {$\boldsymbol{e}_2$};
\draw[->] (x2.north east) to (sum4);
\draw[->] (c2) to (sum4);
\draw[->,densely dotted, out = 80, in = -15] (x2) to (c2);
\draw (sum4) to (split2);
\draw[->] (split2) to (H22);
\draw[->] (H22) to (sum5);
\draw[->] (sum5) to (sum6);
\draw[->] (sum6) to (y2.north west);
\draw[->] (n2) to (sum5);
\draw[->] (e2) to (sum6);
\draw[->,densely dotted,  out = 80, in = -15] (u2) to (e2);
\draw[->] (split2) to (H12);
\draw[->] (H21) to (sum5);
}}
    \caption{Bidirectional IBFD} 
    \label{fig:BidirectionalDigital}
  \end{subfigure}
  \caption{Discrete-time baseband-equivalent system models for IBFD. The quantities $\sqrt{\rho}\boldsymbol{H}$ and $\sqrt{\eta}\boldsymbol{H}$ model signal and interference propagation, respectively, including the effects of propagation-domain and analog-circuit-domain suppression; $\boldsymbol{n}$, $\boldsymbol{c}$, and $\boldsymbol{e}$ represent channel noise, transmitter non-idealities, and receiver non-idealities, respectively; $\boldsymbol{x}$ and $\boldsymbol{s}$ represent intended and actual transmitted signals, respectively; and $\boldsymbol{u}$ and $\boldsymbol{y}$ represent the actual and ADC-measured received signals, respectively.} 
  \label{fig:Digital}
\end{figure}

Focusing on the relay case in \Figref{RelayDigital}, a digital-domain cancellation scheme in the IBFD relay terminal might first compute an estimate $\sqrt{\eta}_{\textsf{r}}\hat{\boldsymbol{H}}_{\textsf{rr}}$ of the (partially-suppressed, direct- and reflected-path) self-interference channel and then use its knowledge of the intended transmit signal $\boldsymbol{x}_{\textsf{r}}$ to form the self-interference estimate $\sqrt{\eta}_{\textsf{r}}\hat{\boldsymbol{H}}_{\textsf{rr}}\boldsymbol{x}_{\textsf{r}}$, which it could subtract from $\boldsymbol{y}_{\textsf{r}}$ \cite{Riihonen11FDMIMO}.
It is important to realize that, in doing so, the effects of channel estimation error, transmit distortion, and receive distortion remain. 
The non-linearity induced by the HPA can be modelled as a polynomial~\cite{Schenk08_RF-Imperfections}. Thus, by estimating the parameters of the polynomial model used to construct the cancellation signal, some of the HPA non-linearities can be suppressed in the digital domain~\cite{Anttila13_Cancellation_PowerAmplifier,Elsayed13_Nonlinear,Bharadia13_fullduplex}. 
In multiple antenna systems, non-linear effects can also be suppressed with a combination of transmit and receive beam-forming~\cite{Antonio:2014}.
By explicitly modeling and accounting for rest of the imperfections too, their deleterious effects on decoding and transmit beamforming can be possibly mitigated.

% Local Variables:
% TeX-master: "FD-tutorial.tex"
% End:

% !TEX root = FD-tutorial.tex

\section{Research Challenges and Opportunities}	\label{sec:research}

In this section, we discuss some of the key research challenges and opportunities that will help to enable the widespread application of IBFD to future wireless networks.

\subsection{Antenna and Circuit Design} \label{sec:radio}

As is evident from the discussions in \secref{techniques}, a significant portion of IBFD self-interference cancellation relies on propagation-domain and analog-circuit-domain techniques. 
As wireless technology continues to spread, new types of mobile devices are being envisioned (e.g., wearable computing innovations like smart-watches, smart-glasses, health bands, and such), many of which have a very small form-factor. 
It is well known, even with half-duplex designs, that antenna-design bears most of the burden when special space- and shape-constraints are imposed on mobile terminals. 
\begin{comment}
 For this reason, propagation-domain self-interference suppression will become even more challenging as device sizes shrink.   
\end{comment}
This underscores the importance of research on circumventing the impact of small-aperture antenna design on IBFD system performance.

While the advancement of digital circuitry has followed Moore's law, and thus benefited from exponential decreases in size and energy consumption, the advancement of analog circuitry has not.
Over the last few decades, some RF processing tasks have successfully migrated to the digital domain (e.g., direct digital conversion) and thus have benefited from Moore's law, although others (e.g., power amplification) have not.
Since the circuit-domain techniques discussed in \secref{analog} often require additional (and often sophisticated) analog circuitry, IBFD terminals face the challenge of balancing self-interference cancellation performance with device cost, real-estate, and power consumption. 
For example, as shown in~\cite{Sahai12-PhaseNoise-Conference, Sahai12-PhaseNoiseJournal,Riihonen:2012asi}, oscillator phase-noise directly impacts the performance of analog domain cancelers. 
Moreover, there exists a complex interplay between different components in the RF chain that affects the gains in spectral-efficiency achievable through IBFD~\cite{Ahmed:2013aa}. 
Thus, an important area of research is the design of low-power, low-cost, small-form-factor antennas, in addition to the analog and digital circuitry needed for self-interference suppression.
In particular, additional gain may be achieved by judiciously designing small-form-factor antennas that facilitate subsequent analog self-interference cancellation.

\subsection{Physical-layer Algorithm Design} \label{sec:phy}

There are numerous important research opportunities for the advancement of IBFD physical-layer communications strategies, such as coding, modulation, power allocation, beamforming, channel estimation, equalization, digital interference cancellation, and decoding.
We outline several below.

A necessary first step for the principled design of any digital communications system is an accurate statistical characterization of the \emph{effective} channel seen by the system.
In the case of IBFD, the effective channel, as seen in the digital domain, subsumes the analog circuitry and antennas as well as the propagation environment, and includes all of their non-idealities as well as the combined effects of any propagation-domain and/or analog-circuit-domain interference suppression strategies that may be employed, as well as ambient noise and external interference.
Measurement-driven experimental studies~\cite{Duarte10-FD-Feasibility,Duarte12-Thesis,Duarte11-FD-Experimental-Characterization,Duarte12_Wifi} have shown that, after self-interference has been (approximately) cancelled, the residual interference can often be well characterized by a Rician distribution whose K-factor is dependent on the amount of direct-path suppression.
Furthermore, these studies have shown that the coherence interval and bandwidth of the residual self-interference depend on the local multipath environment and the terminal mobility, which are yet to be extensively characterized for IBFD terminals. 
Thus, an important area of research is the \term{statistical characterization of the self-interference channel} (with and without cancellation), especially for MIMO systems, where the characterization behaves as a function of array aperture, antenna element polarizations, inter-element distance, and element directionality.
This statistical characterization will become the basis for not only system design but also information-theoretic performance analysis (see for example~\cite{Everett11Allerton}).

Another line of research concerns \term{optimal resource allocation}, and in particular the optimal allocation of limited transmit power over space.
The importance of spatial power allocation was already suggested in \secref{propagation}, where it was noted that alterations of the transmit-antenna directionality pattern for the purpose of self-interference reduction may also have the unintended consequence of reducing the power radiated on the intended receiver's antennas. 
The implication is that IBFD transmitters should be designed not to simply suppress self-interference but rather to optimize the
balance between self-interference suppression and desired-signal radiation.
Fortunately, with electronically steerable antenna arrays, it is possible to perform this optimization on-the-fly, in response to varying environmental conditions. 
Similar opportunities manifest when designing receive-beamforming weights (where array gain becomes the limited resource) or when jointly optimizing the IBFD terminals' transmit- and receive-beamforming weights.
Moreover, when multiple terminals in an IBFD network use steerable antenna arrays, the spatial resource allocation problem extends to the network at large, since in that case multiple terminals can \emph{work together} to spatially separate the desired and self-interference signals seen by each IBFD terminal.
As a concrete example, the distant transmitter may be able to beamform in such a way that, when his signal impinges on the local IBFD terminal, it appears spatially orthogonal to the self-interference experienced by that terminal.
Thus, an important area of research concerns the problem of formulating and solving spatial resource allocation problems in IBFD networks. 
There exists preliminary work on jointly optimizing transmit- and receive-arrays to maximize mutual information in bidirectional \cite{Day12FDRelay} and relay \cite{Day12FDMIMO} MIMO IBFD networks under stylized channel models, but open problems remain in regards to more practical metrics and models.

While the above discussion focused on spatial resource allocation, the time and frequency domains present additional opportunities for optimization.
Take, for example, the case where nearby fast-moving reflections cause the self-interference power to remain extremely high even after all available self-interference reduction strategies have been implemented. 
In this case, forcing the IBFD terminal to operate in true IBFD mode may lead to a total loss of reliable communication.
However, if each signaling epoch was split into two halves, each with an independently adjustable transmit power, the IBFD terminal could effectively fall into a time-domain half-duplex mode and communicate reliably with its partners, albeit at half-duplex rates \cite{Day12FDRelay,Day12FDMIMO}.
Then, when the nearby reflectors leave the scene, the same terminal could easily slip back into a true IBFD operating mode (via uniform temporal power-allocation) and thereby increase its spectral efficiency.
Of course, a similar strategy could be implemented using two non-overlapping frequency bands or two orthogonal code subspaces.
Moreover, in the wideband signal case, where the propagation channels are themselves frequency-selective, it can be seen that the resource allocation problem parameters change as a function of frequency.
Thus, an important area of research concerns the problem of formulating and solving joint space-time-frequency resource allocation problems in IBFD networks. 

Finally, it is important to study the \emph{fundamental performance limits} of IBFD using metrics like Shannon capacity, outage capacity, diversity-multiplexing tradeoff, and network throughput, especially for MIMO systems. 
Since any such analysis will assume particular models for wireless signal propagation and analog/digital circuit non-idealities, the value of each given analysis will ultimately depend on the fidelity of its modeling assumptions. For bidirectional and relay topologies, bounds on the achievable-rate of MIMO IBFD were derived in~\cite{Day12FDRelay,Day12FDMIMO} under the standard isotropic Rayleigh-fading model for wireless signal propagation and a signal-power-dependent AWGN model for the analog/digital circuit non-idealities. 
Furthermore, since IBFD designs can be implemented with differing amounts of radio hardware resources (e.g., shared- versus \separate\ designs as discussed in Section~\ref{sec:history}), capacity gains over half-duplex counterparts can be formulated by either equating RF-circuit resources or number-of-antennas~\cite{Aggarwal:2012aa,Barghi:2012aa}, leading to different conclusions under different use cases.
Clearly, the consideration of other performance metrics, as well as more realistic propagation/circuit models, present important open areas of research.

\subsection{Network Foundations and Protocol Design} \label{sec:net}

A typical wireless terminal of today is but one drop of a sea of networked terminals.  Different networks often have to share the same radio spectrum.  To fully exploit the potentials of IBFD, careful engineering in medium-access control and higher-layer protocols is as important as that in the physical layer.  Research results on the networking aspects of IBFD are emerging and there are abundant research opportunities in this direction.  
Perhaps the biggest research challenge lies in developing a foundation for network design where all or some of the terminals are capable of IBFD operation. 
We discuss several key issues in the following.

IBFD has the potential to significantly increase the overall throughput of a wireless network, beyond simply doubling the spectral efficiency of a point-to-point link.  This is because IBFD removes a major scheduling constraint due to self-collision, so that a terminal may transmit to a second or a group of terminals and simultaneously receive from a third or another group of terminals.  As an example, consider a specific situation in a mobile ad hoc network (MANET), where every terminal wishes to multicast the same data to all of its one-hop neighboring terminals.  Without IBFD, at most one transmission can be successful in each frequency band in each neighborhood, because if multiple terminals simultaneously transmit in the same frequency band, they miss each other's transmission.  IBFD enables all terminals to transmit at the same time, where each terminal receives a superposition of neighbors' transmissions for decoding.  A preliminary study in~\cite{Guo:2010aa} indicates the throughput gain over random access schemes (such as ALOHA) increases without bound as the number of neighbors increases.  Fundamentally, with simultaneous transmissions, each receiver experiences an ergodic multiaccess channel (hence the sum energy is collected), whereas, with intermittent transmissions following a random access protocol, each receiver experiences a nonergodic channel at the frame level, where some transmissions are lost and energy is wasted due to collisions.

From the network perspective, IBFD also allows more concurrent transmissions to be packed in a given area.  An interesting problem is to analyze the throughput of a network of randomly deployed terminals using stochastic geometry~\cite{BacBla09FT, Haenggi12}.  IBFD may have implications on network layer protocols such as routing, as the routing algorithm may not need to try to avoid intersecting routes, which may reduce the length of the route and the overal interference.  In addition, the system throughput may benefit by letting terminals en route jointly process bidirectional flows, perhaps also by leveraging network coding techniques~\cite{fragouli2006network, ZhaLie06Mobicom, KatRah08NET}. 

Admittedly, the information capacity of a general wireless network, with or without IBFD, is widely open.  Perhaps it would be simpler to characterize the capacity advantage due to IBFD in various network scenarios, which would provide useful guidelines to practical design.

Another interesting question concerns the implications of IBFD on practical WiFi-type networks.  An IBFD terminal can be continuously cognitive, even during its own transmission, allowing the terminal to abort its transmission immediately upon collision.  Equally important is improved throughput, e.g., by letting an access point transmit and receive packets simultaneously.
IBFD medium access control schemes for WiFi were proposed in~\cite{achal:fd} and~\cite{Choi:2011m}, with substantial gains reported.  The former includes mechanisms to share random backoff counters and to snoop for full duplex transmission opportunities, whereas the latter greatly alleviates the hidden node problem.

IBFD may also lead to dramatically improved performance in other familiar network scenarios.
In a cognitive radio network, secondary IBFD terminals may cause substantially smaller interference to primary users, and can sustain continuous transmission because it does not need to stop transmission in order to listen to the channel.  
Unlike a half-duplex relay, an IBFD relay can decouple the incoming link and the outgoing link to easy scheduling.  It is also possible to relay the received signal instantaneously with little delay, or relay partially decoded packets before the entire packet is even received.  
IBFD also enables instantaneous feedback between a pair of communicating nodes.  This allows the same transmission rate and higher reliability to be achieved by using shorter variable-length error control codes~\cite{polyanskiy2010channel, chen2013variable-length}, reducing the decoding delay and complexity.
How to fully take advantage of IBFD in such network scenarios is an interesting problem.

Whether IBFD should be transparent to higher layers (MAC layer and above) or be exploited is an important question.   Transparency ensures that protocols designed for half-duplex radios can be used without any changes.  The answer resides in size of the gain achieved by novel protocols taking full advantage of IBFD.

To fully exploit the multiaccess nature of the wireless medium in conjunction with IBFD, one has to overcome the challenge of decoding in presence of co-channel interference or jointly decoding multiple users.  There is a large body of literature on multiuser detection (see, e.g.,~\cite{Verdu98, honig2009advances} and references therein).
Compressed sensing has proven to be a useful technique for multiuser decoding in certain applications~\cite{Zhang:2013aa, Zhang:20XXaa}.
Whether multiple transmissions can and should be synchronized or not is an important issue in practical design.

It is important to note that many of the advantages of full duplexing in a wireless network are retained as long as the terminals are full-duplex at the frame level, even if they are still half-duplex at the symbol level.
In particular, when the power difference between the transmit and the receive signals becomes too large, it may be infeasible or too expensive to cancel self-interference.
A special signaling technique was proposed in~\cite{Guo:2010aa} to achieve virtual IBFD.  
The idea is to introduce off-slots (in time, frequency, or both) within each frame of a terminal's transmissions, through which the terminal can collect useful signals without contamination by self-interference.  This allows a frame to be received (albeit with slot erasures) simultaneously as another frame is being transmitted.
The technique has found applications in neighbor discovery~\cite{Zhang:2013aa}, mutual broadcasting (e.g., for exchanging link and network state information)~\cite{Zhang:20XXaa} as well as ranging and localization~\cite{gan2013distributed}).

% Local Variables:
% TeX-master: "FD-tutorial.tex"
% End:

% !TEX root = FD-tutorial.tex

\section{Conclusions	\label{sec:conclusion}}

In-band full-duplex offers numerous opportunities for increasing the spectral efficiency of wireless networks. The opportunities are understandably accompanied with a number of challenges at all layers, ranging from antenna and circuit design, to the development of theoretical foundations for wireless networks with IBFD terminals. Much work remains to be done, and an inter-disciplinary approach will be essential to meet the numerous challenges ahead.  In short, these are exciting times for wireless researchers and, in turn, the mobile telecommunications industry, who can look forward to more efficient future-generation wireless networks.

\section*{Acknowledgements}

The authors would like to thank the Editor-in-Chief for suggesting to write this review article as a vehicle to introduce in-band full-duplex concepts to the larger research audience. The authors would like to thank Evan Everett, Achaleshwar Sahai, and Jingwen Bai at Rice University for their generous help in preparing the article, which included discussions of prior work, the development of artwork in this paper, and valuable feedback on many drafts.

\appendix

\newcommand{\defn}{\triangleq}
\newcommand{\barst}[1]{|_{#1}}
\newcommand{\PAPR}{\text{\sf PAPR}_x}
\newcommand{\SINR}{\text{\sf SINR}}
\newcommand{\SIR}{\text{\sf SIR}}
\newcommand{\uSINR}{\overline{\text{\sf SINR}}}

This appendix shows that the effective dynamic range of the ADC is approximately $6.02(\Beff - 2)$~dB when the system is designed so that the ADC quantization error is $6.02$~dB (i.e., $1$~bit) below the noise power and the signal's peak-to-average-power ratio is $5$~dB.\footnote{%
The peak-to-average power level (PAPR) varies with signal type, and in particular with symbol constellation, the oversampling factor above the baud rate, the number of carriers (in a multi-carrier system), and the number of antennas (in a MIMO) sytem \cite{Han:2005}. 
For HDSPA waveforms, an often-cited Motorola technical report \cite{Motorola:2004} predicts PAPR between $3$ and $5.9$~dB, while other test models \cite{Braithwaite:2013} estimate a PAPR of $\approx 5.7$.  \label{PAPR}
} 

Suppose that the signal coming into the ADC is
$x=d+s+n$,
where $d$ is the desired signal, $s$ is the self-interference, and $n$ is the ``noise,'' which collects all perturbations beyond self-interference and quantization error.
Then the ADC output will be
$y=x+e$
where, for an ADC with an effective number of bits of $\Beff$, the quantization error $e$ will have variance
$$\sigma_e^2 = \PAPR \sigma_x^2 10^{-6.02 \Beff/10}.$$
Given that each additional bit of $\Beff$ halves the amplitude of the quantization error, we note that the multiplier $-6.02=20\log_{10} \frac{1}{2}$ above plays the role of converting from bits to dB.
Likewise, the multiplier $\PAPR$ above is used to prevent clipping. 
In particular, $\PAPR$ denotes the peak-to-average power level of $x$, defined such that $\sqrt{\PAPR \sigma_x^2}$ is the maximum voltage level at the input to the ADC. 

The signal resulting from perfect digital-domain self-interference cancellation, i.e., $z\defn y-s = d+n+e$, will have a signal to interference-plus-noise ratio (SINR) of $\sigma_d^2/(\sigma_n^2+\sigma_e^2)$ if we treat $e$ as uncorrelated with $n$. 
Assuming that the $\Beff$ has been chosen so that $\sigma_e^2= \sigma_n^2/4$ (i.e., the quantization error power is $6.02$~dB or $1$~bit below the noise power), then the SINR would be 
$$\text{SINR} = 
  \frac{\sigma_d^2}{\sigma_e^2 + \sigma_n^2}=
  \frac{\sigma_d^2}{5\sigma_e^2}=
  \frac{\sigma_d^2}{5 \PAPR (\sigma_d^2 + \sigma_s^2 + \sigma_n^2) 10^{-6.02 \Beff/10}},$$
where $d$, $s$, and $n$ are assumed uncorrelated (so that $\sigma_x^2=\sigma_d^2 + \sigma_s^2 + \sigma_n^2$).
We can then upper-bound the SINR by 
$$\uSINR = \frac{\sigma_d^2}{5 \PAPR \sigma_s^2 10^{-6.02 \Beff/10}}$$
noting that the bound will be tight when $\sigma_s^2 \gg \sigma_d^2 + \sigma_n^2$ (i.e., when the self-interference dominates), which is precisely the case of interest.
Writing the previous expression in dB, we have
$$\uSINR\barst{\rm dB} = \frac{\sigma_d^2}{\sigma_s^2}\barst{\rm dB} 
+ 6.02 \Beff
- \PAPR\barst{\rm dB} 
- 7.0.$$
The above expression shows that, after perfect digital-domain self-interference cancellation, the SINR (in dB) is tightly upper bounded by the pre-ADC signal-to-interference ratio (i.e., $\sigma_d^2/\sigma_s^2\barst{\rm dB}$) plus the dynamic range of the ADC (i.e., $6.02 \Beff$~dB) plus a penalty factor of $- \PAPR\barst{\rm dB} - 7$ dB.
If we use\footnote{%
See footnote \ref{PAPR}.}
$\PAPR=5$~dB, then the penalty factor becomes $-12$~dB, which amounts to a loss of $2$~bits relative to the $\Beff$.

\bibliographystyle{ieeetr}
\bibliography{IEEEabrv,bibs/blissFD,bibs/ashu-FD,bibs/ashu-general,bibs/FDbib,bibs/guo,bibs/rw}

\end{document}